\documentclass[twocolumn,showpacs,aps,prb,amsmath,amssymb]{revtex4}

\usepackage{graphicx}

\begin{document}

\title{An exact algorithm for spin correlation functions of the two dimensional
 $\pm J$ Ising spin glass in the ground state}

\author{J. Poulter}
\affiliation{Department of Mathematics, Faculty of Science,
Mahidol University, Rama 6 Road, Bangkok 10400, Thailand}

\author{J.A. Blackman}
\affiliation{Department of Physics, University of Reading, P.O.
Box 220, Reading RG6 6AF, UK}


\begin{abstract}

We introduce an exact algorithm for the computation of spin
correlation functions for the two dimensional $\pm J$ Ising spin
glass in the ground state. Unlike with the transfer matrix method,
there is no particular restriction on the shape of the lattice
sample, and unlike Monte Carlo based methods it avoids
extrapolation from finite temperatures. The computational
requirements depend only on the number and distribution of
frustrated plaquettes.

\end{abstract}


\maketitle

\section{Introduction}

Due to its comparative simplicity, the short range bimodal $\pm J$
Ising system is a widely studied model of a spin glass. The
Hamiltonian is of the form proposed by Edwards and Anderson \cite
{EA75}

\begin{equation}
H=-\sum_{<ij>}J_{ij}\sigma_{i}\sigma_{j}
\label{e:EAH}
\end{equation}
where the nearest-neighbor exchange interactions $J_{ij}$ are
quenched random variables of fixed magnitude but random sign.
These bonds are negative with a probability $p\in [0,0.5]$ for the
square lattice. The canonical model with $p=0.5$ on the square
lattice has been studied the most and it is well accepted that
spin glass behavior occurs at zero temperature \cite
{KR97,KS00,HY01,H01,SP03,KL04}, and persists down to a critical
probability $p_{c}$ of about $0.11$ \cite
{K77,MB80,ON87,KO90,UO91,BGP98,RQS99,HPP01,N01,MC02,NN02,AH04}.

One of the compelling features of the 2-dimensional Ising model is
the special property of allowing exact solutions - at least for
finite systems in the absence of a magnetic field. A number of
authors have taken this approach in various guises
\cite{BGP98,MC02,B82,BP91,PB01,SK93,BMRU82,DM91,PA99}. One
advantage of doing so is that it provides a direct access to the
ground state properties without the need to extrapolate from
finite temperatures as is necessary, for example, in Monte Carlo
based methods.

Good agreement for the values of the ground state energy and
entropy is obtained by the various workers. However, to the
present, there has been very little development in extending the
methodology to a direct calculation of the spin correlations at
zero temperature, and most of the results have come from
extrapolations from finite temperatures.

The present authors developed an approach
\cite{BGP98,B82,BP91,PB01}, based on the pfaffian matrix, that
appears to capture the essence of the physics of the $\pm J$
system. The algorithm enables certain quantities such as the
ground state free energy and entropy to be calculated exactly for
very large lattices. The objective of the current paper is to
extend that methodology to give direct access to the zero
temperature correlation functions.

Spin correlations for the bimodal Ising spin glass are expected to decay
algebraically according to

\begin{equation}
[<\sigma_{0}\sigma_{R}>^{2}]_{av}\sim R^{-\eta}
\label{e:SS1}
\end{equation}
at the critical temperature. In practice, finite size effects
dictate that the spin correlations will decay exponentially

\begin{equation}
[<\sigma_{0}\sigma_{R}>^{2}]_{av}\sim R^{-\eta} \exp(-R/\xi)
\label{e:SS2}
\end{equation}
where the correlation length $\xi$ is expected to be proportional
to the system size if the latter is large enough.

To our knowledge the only direct computations of spin correlation functions
in the ground state are those of Ozeki \cite {Ozeki}. This work used a
numerical transfer matrix method with long thin samples of circumference $L$
wrapped around a cylinder of length $9L$ with open ends. The maximum possible
circumference for this study was only $L=12$, a consequence of the transfer
matrix computational requirements scaling exponentially with $L$.

All other attempts to study spin correlations in the ground state have
involved extrapolation from finite temperature. Monte Carlo techniques
have been employed to obtain results for low temperatures, for example
$0.86J$ \cite {McMillan} and $0.63J$ \cite {WS88}. However, it has never been
clear just how reliable these extrapolations are.

The transfer matrix method has also been used to obtain spin
correlations functions at finite temperature \cite
{RQS99,HPP01,QS03}, although the sample shape restrictions are
severe with cylindrical circumference $L\leq 20$. A better
approach is probably the network model \cite {MC02} where the
computational requirements scale as $L^{3}$, not exponentially.
Nevertheless, although larger values of $L$ are feasible, the zero
temperature limit is inaccessible.

The algorithm we employ depends only on the number and
distribution of frustrated plaquettes. This means that there is no
need for the cylindrical circumference to be especially small and
the two lattice dimensions can be treated on an essentially equal
footing. Although this algorithm is related in some formalities to
the network model, it is especially designed to operate in the
ground state.

The method is also fully gauge invariant. This means that it is
well suited for the determination of gauge invariant quantities.
In this context, a transformation of disorder is gauge invariant
if the number and distribution of frustrated plaquettes is
unchanged. Examples of gauge invariant quantities are energy,
entropy and the squares of correlation functions. In contrast,
matching algorithms \cite {BMRU82,DM91,PA99}, although more
efficient, cannot determine more than the energy.

The formalism is developed in Section 2, and is then used in
Section 3 for the evaluation of $\eta$ for the canonical $\pm J$
model.

\section{Formalism}

The planar Ising model has long been analytically accessible since it can
be mapped onto a system of non-interacting fermions. This mapping can take
various forms, including the transfer matrix and combinatorial methods.
The combinatorial, or Pfaffian, method \cite {GH64} is particularly well
suited to the study of disordered planar Ising systems. Essentially, each
lattice site is decorated with four fermions which have interactions across
bonds as well as intrasite interactions.

The partition function for a disordered planar Ising model can be expressed
in the form

\begin{equation}
Z=2^{N}[\prod_{<ij>} \cosh (J_{ij}/kT)] \hspace {2 mm} (\det D)^{1/2}
\label{e:Z}
\end{equation}
where the product is over all nearest neighbor bonds $J_{ij}$ on
the $N$ site lattice and $D$ is a skew-symmetric matrix. The
square root of the determinant of $D$ is the major feature and is
precisely the Pfaffian \cite {GH64}. It is also proportional to
the trace over all closed lattice polygons, that is

\begin{equation}
Tr \prod_{<ij>}(1+t_{ij}\sigma_{i}\sigma_{j})=2^{-N}(\det D)^{1/2}
\label{e:trace}
\end{equation}
where $t_{ij}=\tanh(J_{ij}/kT)$.

The correlation functions can be expressed within the same
formalism as a reciprocal defect problem \cite {B82,GH64}. We
choose a path between a pair of spins and replace $t_{ij}$ with
$t_{ij}^{-1}$ for bonds on that path. It may be noted here that
the path can also be disjoint, in which case the formalism would
give a correlation function for four or more spins. In terms of
determinants, the correlation function can be easily expressed as

\begin{equation}
<\sigma_{0}\sigma_{R}>^{2}=\frac{\det C}{\det D}
\prod_{path}t_{ij}^{2} \label{e:ss}
\end{equation}
where the matrix $C$ is the same as $D$ except for the reciprocal
defects.

The calculation of the partition function for the disordered Ising
model has been described before \cite {BP91}. The key points are
as follows. At zero temperature, $D$ is a singular matrix with
zero eigenvalues equal in number to the number of frustrated
plaquettes. These eigenvalues (which occur in pairs) approach zero
as some power of $\exp(-2J/kT)$

\begin{equation}
\epsilon=\pm \frac{1}{2}X \exp(-2Jr/kT)
 \label{e:eps}
\end{equation}
where $r$ is an integer. The ground state energy and entropy can
be expressed exactly as

\begin{equation}
F=-2J+2J \sum_{d} r_d
 \label{e:deltaf}
\end{equation}

\begin{equation}
S=k \sum_{d} \ln X_d
 \label{e:s}
\end{equation}
The sums are over all zero eigenstates. An algorithm
based on degenerate state perturbation theory was developed \cite
{BP91} to evaluate exactly the $r$ and $X$, and hence the ground
state energy and entropy. In this paper we show how to extend this
algorithm to the correlation functions as well.

\begin{figure}[h]
\includegraphics[trim=80 620 200 100,width=8.5cm]{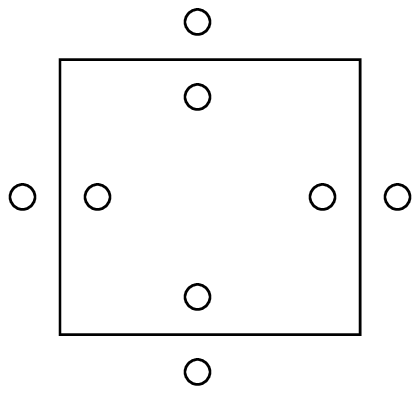}
\caption{\label{f:Fig1} A plaquette showing how lattice fermions
are associated with bonds and plaquettes.}
\end{figure}

It was found \cite {BP91} to be convenient to transform the
lattice fermions from the sites to the bonds. In this way, we can
associate two fermions with each bond and, on the square lattice,
four fermions with each plaquette. Figure \ref{f:Fig1} provides an
illustration. A simple generalization to another planar lattice
has proved straightforward \cite {PB01}.

For the $\pm J$ model in particular we write

\begin{equation}
D=D_{0}+\delta D_{1}
\label{e:D}
\end{equation}
where $\delta=1-t$ with $t=\tanh(J/kT)$. This equation is exact
and $D_{1}$ characterizes a perturbation away from the singular
matrix $D_{0}$ and has nonzero matrix elements only across bonds.
The singularities of $D_{0}$ arise due to the frustration. For
each frustrated plaquette, $D_{0}$ has a zero eigenvalue with an
eigenvector localized on the four associated fermions. To
determine the ground state properties, it is necessary to
determine the defect eigenvalues of $D$, that is those eigenvalues
that are zero at zero temperature. First $D_{1}$ is diagonalized
in the basis of the defect eigenvectors localized on the
frustrated plaquettes. The second order calculation then involves
diagonalizing

\begin{equation}
D_{2}=D_{1}g_{c}D_{1}
\label{e:D2}
\end{equation}
in the basis of the eigenvectors corresponding to zero eigenvalues at first
order. This process is continued order by order where, for $n>2$,

\begin{equation}
D_{n}=D_{n-1}(1+G_{n-2}D_{n-2})\cdots(1+G_{1}D_{1})g_{c}D_{1}
\label{e:DSPT}
\end{equation}
until no zero eigenvalues remain. In these equations,
$g_{c}$ is the continuum propagator and $G_{n}$ is the propagator
for eigenstates determined at order $n$ \cite {BP91}.

The matrix $C$ is given by

\begin{equation}
C=D_{0}+\delta(D_{1}+2V)+(\delta^{2}+\delta^{3}+\cdots)V
\label{e:C}
\end{equation}
where $V=-D_{1}$ for matrix elements across a path bond. All other elements
of $V$ are zero. This equation is easily derived by expanding $t^{-1}-t$ in
powers of $\delta$. Since $t^{2}=1$ in the ground state, we can now state our
main task as the computation of the limit

\begin{equation}
[<\sigma_{0}\sigma_{R}>^2]_{T=0}=\lim_{\delta\rightarrow 0} \frac{\det C}{\det D}
\label{e:limit}
\end{equation}

To achieve this goal efficiently, we have discovered a remarkable property
of the continuum propagator. It can be written as the sum of two terms

\begin{equation}
g_{c}=g_{c1}+g_{c2}
\label{e:gc}
\end{equation}
where $g_{c1}$ is $4 \times 4$ block diagonal in the four fermions
associated within a plaquette and $g_{c2}$ is $2 \times 2$ block
diagonal in the two fermions associated with a bond. The physical
relevance of this decomposition can be realized in that $g_{c2}$
should not play any role in the ground state, being only an issue
for excited states. Both $g_{c2}$ and $D_{1}$ are bond diagonal,
meaning that $g_{c2}D_{1}$ costs energy while taking us nowhere.
We can now state an important theorem concerning the determination
of $\det C$.

\vspace{5 mm} \noindent{\bf Theorem.} {\it In the ground state
limit, the solution for} $\det C$ {\it using}
\begin{align}
C&=D_{0}+\delta(D_{1}+2V)+(\delta^{2}+\delta^{3}+\cdots)V \nonumber \\
g_{c}&=g_{c1}+g_{c2} \label{e:SYS1}
\end{align}
{\it is exactly the same as the solution using}
\begin{align}
C&=D_{0}+\delta(D_{1}+2V) \nonumber \\
g_{c}&=g_{c1} \label{e:SYS2}
\end{align}
\vspace{1 mm}

Essentially, the effects of $g_{c2}$ exactly cancel the awkward
nonlinear terms in $\delta$. We can simply rerun the perturbation
theory for $D_{1}+2V$ instead of $D_{1}$, noting that this simply
requires a sign change across path bonds. Further, there is an
obvious corollary that says that $g_{c2}$ can be simply ignored
while computing $\det D$. The continuum propagator is really just
$g_{c1}$ and is localized inside plaquettes. The physical sense of
this is clear. A proof of the theorem is given in the appendix.

In \cite {BP91} it was proven that $\det D$ is proportional to the
ground state degeneracy (Equation (\ref{e:s}) is an equivalent
statement). We can generalize this by considering $\det C$ as
proportional to an effective degeneracy for the reciprocal defect
system. In fact we can usually write

\begin{equation}
<\sigma_{0}\sigma_{R}>^{2}=\exp(2(S_{C}(R)-S_{D}))
\label{e:entropy}
\end{equation}
where $S_{D}$ is the system entropy and $S_{C}$ is the analog for
the reciprocal defect system. This result applies unless the
effective energy of the reciprocal defect system is less than the
actual energy in which case the correlation function is zero. We
are not aware of this result appearing elsewhere in the
literature. As a function of system size $L$, the entropy is
expected to vary as \cite {BGP98}

\begin{equation}
S_{D}(L)=S_{D1}+S_{D2}/L
\label{e:FSS}
\end{equation}
with $S_{D2}>0$ for $p>p_{c}$ and $S_{D2}<0$ otherwise. This same
behavior is also expected with cylindrical winding, that is
periodic boundary conditions applied in one dimension. From this,
we can of course reasonably expect that the first cumulant
approximation

\begin{equation}
<\sigma_{0}\sigma_{R}>^{2} \sim A(R)\exp(-B(R)/L)
\label{e:expss}
\end{equation}
is valid if $L$ is sufficiently large. This is of course in accordance
with finite size scaling theory, where it is expected that $A(R)\sim R^{-\eta}$
and $B(R)\sim R$ if $R$ is sufficiently large. The nature of higher order
corrections with cylindrical winding is unclear \cite {CHK04}.

\section{Results}

\begin{figure}[h]
\includegraphics[angle=-90,trim=80 60 40 100,width=8.5cm]{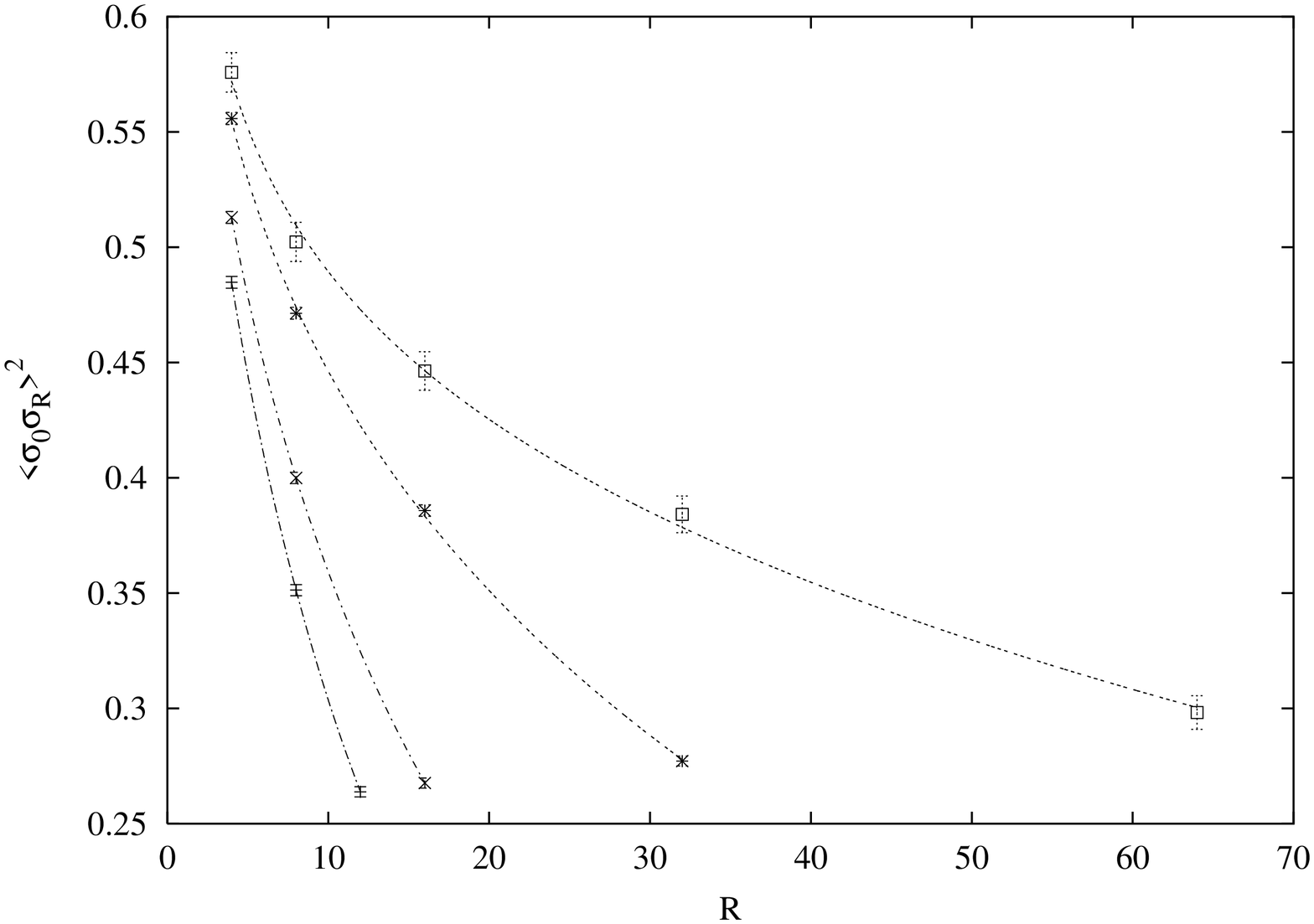}
\caption{\label{f:Fig2} The square of the spin correlation
function on $9L \times L$ lattices with $L=12$ (pluses), $16$
(crosses), $32$ (stars) and $64$ (squares).}
\end{figure}

We have first estimated correlation functions with paths of length
$R$, in units of the lattice spacing, parallel to the axes of long
thin cylinders of dimensions $9L \times L$ with $L=12$, $16$, $32$
and $64$. Ozeki \cite {Ozeki} used similar geometry, but the
largest size treated was $L=12$. The results are displayed in
figure \ref{f:Fig2}. The error bars indicate uncertainties equal
to two standard deviations, that is a $95.4$ per cent confidence
interval. The averages are over $10^{5}$ random samples, except
for $L=64$ where only $10^{4}$ samples were obtained. The
uncertainties do not depend strongly on either $L$ or $R$, only on
the number of samples. With $10^{5}$ samples, they all sit in the
range $0.0021$ to $0.0026$. The corresponding interval with
$10^{4}$ samples is $0.0073$ to $0.0086$, about a factor of $\surd
10$ larger. The curves are fits of the data to the form

\begin{equation}
<\sigma_{0} \sigma_{R}>^{2} = A R^{-\eta} \exp(-R/\xi)
\label{e:scale}
\end{equation}
These $\chi ^{2}$ nonlinear fits were done using the
Levenberg-Marquardt method \cite {numrecip} with the statistical
uncertainties incorporated.

The coefficient $A$ and the exponent $\eta $ are both
approximately independent of $L$. The values of $A$ are $0.73(4)$,
$0.72(2)$, $0.72(1)$ and $0.71(3)$ respectively for $L=12$, $16$,
$32$ and $64$. The corresponding values of $\eta$ are $0.12(6)$,
$0.14(3)$, $0.15(1)$ and $0.14(2)$. In comparison, the correlation
length $\xi$ does not vary like $L$ as would be expected from
finite size scaling theory. In fact we find that it varies more
like $L^{\frac{3}{2}}$. The respective values of $\xi /
L^{\frac{3}{2}}$ are $0.41(6)$, $0.41(4)$, $0.40(3)$ and
$0.47(10)$. We also find that finite size corrections to the
entropy for the $9L \times L$ system scale like
$L^{-\frac{3}{2}}$, whereas scaling as equation (\ref{e:FSS})
applies with the $L \times L$ lattice. Clearly the correlation
functions are intimately associated with the entropy as is
highlighted in equation (\ref{e:entropy}). Interestingly Lukic et
al \cite {LGM04} also observe finite size corrections to the
ground state entropy scaling like $L^{-\frac{3}{2}}$, albeit for a
system with different geometry.

\begin{figure}[h]
\includegraphics[angle=-90,trim=80 60 40 100,width=8.5cm]{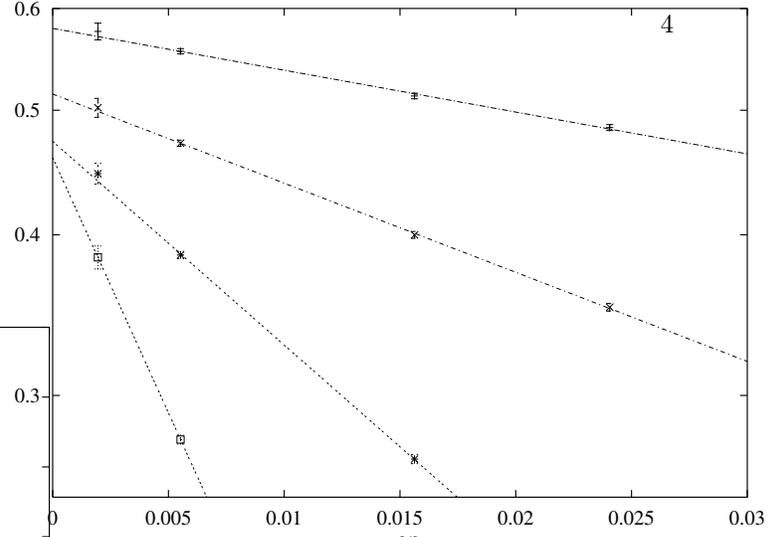}
\caption{\label{f:Fig3} As figure \ref{f:Fig2} with the data
plotted against $L^{-\frac{3}{2}}$ for $R=4$ (pluses), $8$
(crosses), $16$ (stars) and $32$ (squares).}
\end{figure}

Figure \ref{f:Fig3} shows the same data plotted against
$L^{-\frac{3}{2}}$ for $R=4$, $8$, $16$ and $32$. The $\chi ^{2}$
fits are to the function $\alpha (R) \exp (-\beta (R)
L^{-\frac{3}{2}})$ and were first done with linearized data before
polishing with nonlinear fits to obtain uncertainties for the
fitting parameters $\alpha (R)$ and $\beta (R)$. A consequent fit
of $\alpha (R)$ to a power function $R^{-\eta}$ gives $\eta =
0.14(1)$. We have also tried to fit the data to the form of
equation (\ref{e:expss}). The result is certainly inferior with
considerably larger $\chi ^{2}$ values for $R=8$ and $16$ in
particular, although $B(R)$ does follow $R$ rather roughly. In
detail, for $R=4$, $8$, $16$ and $32$, the respective values of
$B(R)/R$ are $0.65(6)$, $0.69(4)$, $0.72(4)$ and $0.65(9)$.

\begin{figure}[h]
\includegraphics[angle=-90,trim=80 60 40 100,width=8.5cm]{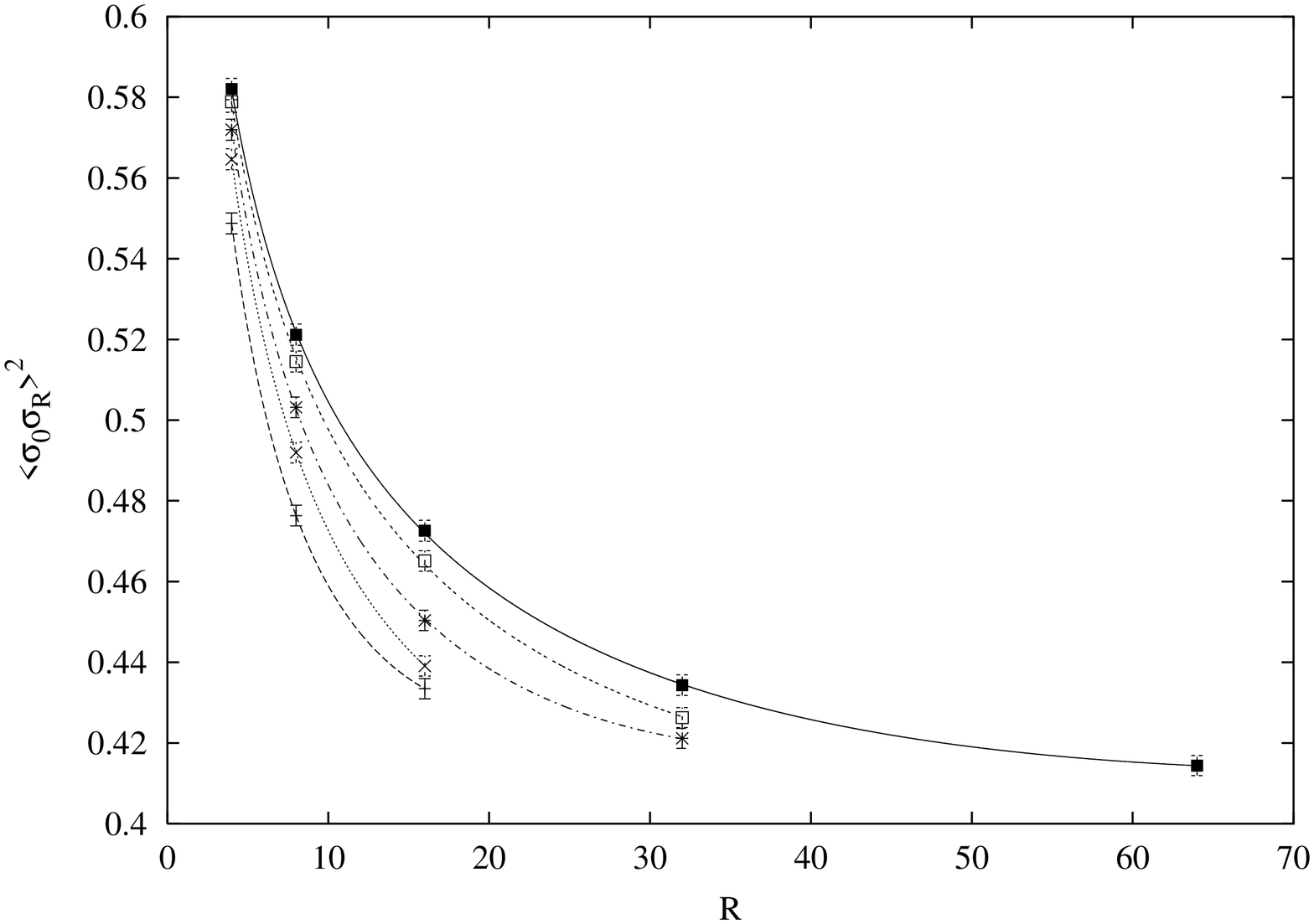}
\caption{\label{f:Fig4} The square of the spin correlation
function on $L \times L$ lattices with $L=32$ (pluses), $48$
(crosses), $64$ (stars), $96$ (squares) and $128$ (filled
squares).}
\end{figure}

We have also performed estimations of correlation functions on $L
\times L$ lattices. The lattices are cylindrically wound as before
but the reciprocal defect path is around the circumference of the
cylinder, as far as possible from the nested boundaries. Figure
\ref{f:Fig4} shows the results for a range of values of $L$ up to
128. The number of random samples taken was $10^{5}$. The curves
are a fit to the form of equation (\ref{e:scale}) and serve only
as a guide to the eye. The statistical uncertainties for our
estimates, with $10^{5}$ random samples, on $L \times L$ lattices
are in the interval $0.0021$ to $0.0026$, the same as for the $9L
\times L$ lattices.

\begin{figure}[h]
\includegraphics[angle=-90,trim=80 60 40 100,width=8.5cm]{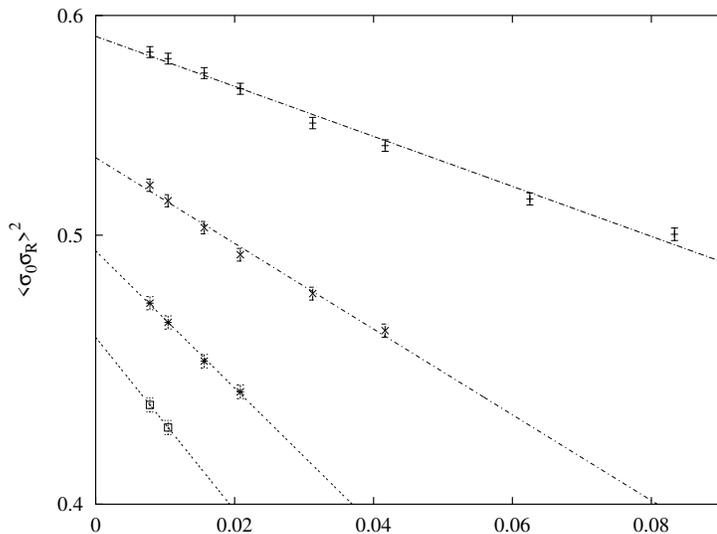}
\caption{\label{f:Fig5} The square of the spin correlation
function on $L \times L$ lattices with $R=4$ (pluses), $8$
(crosses), $16$ (stars) and $32$ (squares).}
\end{figure}

Figure \ref{f:Fig5} shows data plotted against $L^{-1}$. The fits
are to the form of equation (\ref{e:expss}). Only data for which
$L>2R$ is used since it is observed that data for smaller $L$ sits
significantly above the fit. This is most probably a consequence
of the finite size effect of the cylindrical circumference. A
further fit of $A(R)$ to $R^{-\eta}$ reveals $\eta =0.13(1)$ which
is in reasonable agreement with the values found above.

\section{Discussion}

Earlier work on the exact calculation of the energy and entropy of
the 2-dimensional $\pm J$ spin glass has been extended to cover
spin correlations in the ground state. The computational
requirements of the algorithm depend only on the number and
distribution of frustrated plaquettes, and this provides the
possibility of studying quite large samples. The big advantage of
the approach is that there is no need to extrapolate from finite
temperatures as in Monte Carlo methods.

Our best estimate of the exponent is $\eta = 0.14$. We regard the
cylindrical winding as the most reliable configuration for
calculation since it provides one long dimension. This value for
$\eta$ emerges from the calculation for the largest value of $L$,
and for the composite analysis in Figure \ref{f:Fig3}. The
calculation on the $L \times L$ lattice also yields a similar
figure.

A wide range of values of $\eta$ have been reported in the
literature over the years. A significant number of these favor a
value around $0.2$, somewhat higher than ours, but most of these
use extrapolated Monte Carlo data rather than focusing directly on
the ground state. Interestingly a recent calculation \cite{KL04}
reports $\eta = 0.138\pm 0.005$ in agreement with the current
calculation.

\appendix
\section{}

Although the central theorem of this article makes good physical
sense, it can also be proven with mathematical rigor. This
appendix outlines a proof. For present purposes, all matrices are
defined as pure imaginary Hermitian, that is with real
eigenvalues.

The main issue is finding the eigenvalues of

\begin{equation}
C=D_{0}+\delta (D_{1}+2V) + (\delta^{2}+\delta^{3}+\cdots)V
\label{e:A1}
\end{equation}
The eigenvalue equation is formally written

\begin{equation}
C \mid \Psi \rangle = \lambda \mid \Psi \rangle
\label{e:A2}
\end{equation}
and $\lambda$ and $\mid\Psi\rangle$ can be expanded in powers of $\delta$

\begin{equation}
\lambda=\delta^{n} \epsilon_{n}^{j} + O(\delta^{n+1})
\label{e:A3}
\end{equation}
\begin{equation}
\mid\Psi\rangle = \mid\Psi_{0}\rangle + \delta \mid\Psi_{1}\rangle + \cdots
\label{e:A4}
\end{equation}
where the eigenvalue is the $j$th defect eigenvalue of order $n$ with
$n>0$ and $\mid\Psi_{0}\rangle=\mid n,j\rangle$ is the leading term
for the corresponding eigenvector, that is $D_{0} \mid n,j\rangle = 0$.
The notation here is in keeping with \cite {BP91}.

For $1 \leq m \leq n$, equating powers of $\delta$, we obtain

\begin{equation}
D_{0}\mid\Psi_{m}\rangle+ C_{1}\mid\Psi_{m-1}\rangle+
V\sum_{p=2}^{m}\mid\Psi_{m-p}\rangle
=\delta_{mn}\epsilon_{n}^{j}\mid n,j\rangle
\label{e:A5}
\end{equation}
where $C_{1}=D_{1}+2V$ and the sum vanishes for $m<2$.
>From this it follows that

\begin{equation}
\langle r,i \mid C_{1} \mid \Psi_{m-1}\rangle
+\sum_{p=2}^{m}\langle r,i\mid V\mid\Psi_{m-p}\rangle
=\delta_{mn} \delta_{rn} \delta_{ij} \epsilon_{n}^{j}
\label{e:A6}
\end{equation}
Then, for $m=n=r=1$ we now have

\begin{equation}
\langle 1,i\mid C_{1}\mid 1,j\rangle = \delta_{ij} \epsilon_{1}^{j}
\label{e:A7}
\end{equation}
and the first order states are determined by diagonalization of
$C_{1}$ in the defect basis set, that is the localized vectors
associated with the frustrated plaquettes.

Following a development similar to that given for the matrix $D$ in \cite
{BP91} we can show that

\begin{eqnarray}
\mid\Psi_{m}\rangle = g_{c} C_{1} \mid \Psi_{m-1}\rangle + g_{c} V
\sum_{p=2}^{m}\mid\Psi_{m-p}\rangle \nonumber\\ - \delta_{mn}
g_{c} \epsilon_{n}^{j}\mid n,j\rangle + \sum_{r,i}\mid r,i\rangle
Z_{r}^{i} \label{e:A8}
\end{eqnarray}
where $Z_{r}^{i}$ is an undetermined coefficient.

For $1\leq m\leq n-1$, we now use equations (\ref{e:A6}) and (\ref{e:A8}) to proceed
to the second order problem. As stated in the main text, we write
$g_{c}=g_{c1}+g_{c2}$. Further, it is easy to demonstrate that
$g_{c2}D_{1}=\frac{1}{2}$, $Vg_{c2}V=-\frac{1}{2}V$,
$C_{1}g_{c2}V=-\frac{1}{2}V$ and $C_{1}g_{c2}C_{1}+V=\frac{1}{2}C_{1}$.
Then, defining $C_{2}=C_{1}g_{c1}C_{1}$, it can be shown that, for $r>1$,

\begin{eqnarray}
\langle r,i\mid C_{2} \mid\Psi_{m-1}\rangle +
\sum_{p=2}^{m}\langle r,i\mid C_{1}g_{c1}V\mid\Psi_{m-p}\rangle
\nonumber\\= \delta_{m,n-1}\delta_{rn}\delta_{ij}\epsilon_{n}^{j}
\label{e:A9}
\end{eqnarray}
and we can determine the second order states using

\begin{equation}
\langle 2,i\mid C_{2} \mid 2,j\rangle = \delta_{ij} \epsilon_{2}^{j}
\label{e:A10}
\end{equation}
Also, with $r=1$, we can now find the coefficients

\begin{equation}
Z_{1}^{i} = -\frac{1}{\epsilon_{1}^{i}}\langle 1,i\mid C_{2}\mid\Psi_{m-1}\rangle
            -\frac{1}{\epsilon_{1}^{i}}\sum_{p=2}^{m}
            \langle 1,i\mid C_{1}g_{c1}V\mid\Psi_{m-p}\rangle
\label{e:A11}
\end{equation}
and, with the definition

\begin{equation}
F_{r} = - \sum_{i} \mid r,i\rangle \frac{1}{\epsilon_{r}^{i}}
\langle r,i\mid , \label{e:A12}
\end{equation}
we find that

\begin{align}
\mid\Psi_{m}\rangle& = \left((1+F_{1}C_{1})g_{c1}+g_{c2}\right)
\nonumber \\ \times& \left(C_{1} \mid\Psi_{m-1}\rangle + V
\sum_{p=2}^{m} \mid\Psi_{m-p}\rangle\right)
         + \sum_{r>1,i} \mid r,i\rangle Z_{r}^{i}
\label{e:A13}
\end{align}
Note that we are using $F_{r}$ for the propagators of the system with
reciprocal defects to avoid confusion with the $G_{r}$ used in \cite {BP91}
for the diagonalisation of $D$.

Next we consider $1\leq m\leq n-2$ and work with equations (\ref{e:A9}) and
(\ref{e:A13}) as well as the results
$C_{2}g_{c2}V=-\frac{1}{2}C_{1}g_{c1}V$ and
$C_{2}g_{c2}C_{1}=\frac{1}{2}C_{2} - C_{1}g_{c1}V$.
We also define $C_{3}=C_{2}(1+F_{1}C_{1})g_{c1}C_{1}$.
Then, for $r>2$, we can arrive at

\begin{align}
\langle r,i\mid C_{3} \mid\Psi_{m-1}\rangle +& \sum_{p=2}^{m}
\langle r,i \mid C_{2}(1+F_{1}C_{1})g_{c1}V \mid\Psi_{m-p}\rangle
\nonumber \\ =& \delta_{m,n-2} \delta_{rn} \delta_{ij}
\epsilon_{n}^{j}  \label{e:A14}
\end{align}
Clearly the third order problem is solved by diagonalizing
$C_{3}$. Furthermore, for $r=2$, it is possible to determine the
coefficients $Z_{2}^{i}$ and we find that

\begin{align}
\mid\Psi_{m}\rangle& =
\left((1+F_{2}C_{2})(1+F_{1}C_{1})g_{c1}+g_{c2}\right) \nonumber
\\ \times& \left(C_{1} \mid\Psi_{m-1}\rangle
   + V \sum_{p=2}^{m} \mid\Psi_{m-p}\rangle \right)
   + \sum_{r>2,i} \mid r,i\rangle Z_{r}^{i}
\label{e:A15}
\end{align}

To prove the main result, we can use induction. First we define, for
$k\geq3$,

\begin{equation}
C_{k}=C_{k-1}(1+F_{k-2}C_{k-2})\cdots(1+F_{1}C_{1})g_{c1}C_{1}
\label{e:A16}
\end{equation}
and, for $1\leq m\leq n-k+1$ and $r\geq k-1$, introduce two assumptions.
First,

\begin{align}
\langle r,i \mid C_{k}& \mid\Psi_{m-1}\rangle + \nonumber  \\
\sum_{p=2}^{m} \langle r,i& \mid C_{k-1}(1+F_{k-2}C_{k-2})\cdots
(1+F_{1}C_{1})g_{c1}V \mid\Psi_{m-p} \rangle   \nonumber  \\
&= \delta_{m,n-k+1} \delta_{rn} \delta_{ij} \epsilon_{n}^{j}
\label{e:A17}
\end{align}
and, second

\begin{eqnarray}
\mid\Psi_{m}\rangle =
\left((1+F_{k-1}C_{k-1})\cdots(1+F_{1}C_{1})g_{c1}+g_{c2}\right)
\nonumber \\ \times \left(C_{1}\mid\Psi_{m-1}\rangle +V
\sum_{p=2}^{m} \mid\Psi_{m-p}\rangle \right) + \sum_{r\geq k,i}
\mid r,i\rangle Z_{r}^{i} \nonumber \\ \label{e:A18}
\end{eqnarray}

We can note that, comparing with equations (\ref{e:A14}) and (\ref{e:A15}),
these assumptions are both true for $k=3$. Also, with the
definition (\ref{e:A16}), we can easily show that

\begin{equation}
C_{k}g_{c2}V = -\frac{1}{2}C_{k-1}(1+F_{k-2}C_{k-2})\cdots
                    (1+F_{1}C_{1})g_{c1}V
\label{e:A19}
\end{equation}
and

\begin{equation}
C_{k}g_{c2}C_{1}=\frac{1}{2}C_{k}
 - C_{k-1}(1+F_{k-2}C_{k-2})\cdots(1+F_{1}C_{1})g_{c1}V
\label{e:A20}
\end{equation}
Now, for $1\leq m\leq n-k$ and $r\geq k$, we can use the assumptions
(\ref{e:A17}) and (\ref{e:A18}) with
(\ref{e:A19}) and (\ref{e:A20}) to prove two results. First, for $r>k$,

\begin{align}
\langle r,i \mid C_{k+1} \mid&\Psi_{m-1}\rangle +\nonumber  \\
\sum_{p=2}^{m}\langle r,i \mid C_{k}&(1+F_{k-1}C_{k-1})\cdots
(1+F_{1}C_{1})g_{c1}V \mid\Psi_{m-p}\rangle \nonumber   \\
 &= \delta_{m,n-k} \delta_{rn} \delta_{ij} \epsilon_{n}^{j}
\label{e:A21}
\end{align}
and, second, after determining the coefficients $Z_{k}^{i}$,

\begin{align}
\mid\Psi_{m}\rangle =&
\left((1+F_{k}C_{k})\cdots(1+F_{1}C_{1})g_{c1}+g_{c2}\right)
\nonumber\\ \times& \left(C_{1} \mid\Psi_{m-1}\rangle
+ V \sum_{p=2}^{m}\mid\Psi_{m-p}\rangle \right) \nonumber \\
+& \sum_{r>k,i} \mid r,i \rangle Z_{r}^{i} \label{e:A22}
\end{align}

This completes the proof of the theorem. Clearly  the assumptions (\ref{e:A17})
and (\ref{e:A18}) imply the results (\ref{e:A21}) and (\ref{e:A22}).
The eigenvalues of $C$ are determined by diagonalisations of the $C_{r}$
which are defined independently of $g_{c2}$.

\end{document}